\documentclass[twoside,twocolumn,10pt]{article}

\newcommand{\rev} [1]{{\textcolor{black}{#1}}}

\usepackage{wscg}           
\usepackage{hyperref}
\usepackage{algorithm}
\usepackage{algpseudocode}
\usepackage{booktabs}
\usepackage{subcaption}

\RequirePackage{ifpdf}
\ifpdf
 \RequirePackage[pdftex]{graphicx}
 \RequirePackage[pdftex]{color}
\else
 \RequirePackage[dvips,draft]{graphicx}
 \RequirePackage[dvips]{color}
\fi

\usepackage{nopageno}       
\usepackage[switch]{lineno}

\title{A Robust Approach to Detect Intersections between Triangles with Different Numerical Representations } 

\author{
\parbox{0.4\textwidth}{\centering
Luca Garau\\[1mm]
University of Cagliari\\
Via Ospedale 72\\
09124, Cagliari, Italy\\[1mm]
luca.garau@unica.it
}
\hspace{0.1\textwidth}
\parbox{0.4\textwidth}{\centering
Gianmarco Cherchi\\[1mm]
University of Cagliari\\
Via Ospedale 72\\
09124, Cagliari, Italy\\[1mm]
g.cherchi@unica.it
}
}

\usepackage{url}
\makeatletter
\def\Uslash{\mathbin{\mathchar`\/}\@ifnextchar{/}{\kern-.15em}{}}
\g@addto@macro\UrlSpecials{\do \/ {\Uslash}}
\def\Ucolon{\mathbin{\mathchar`:}\@ifnextchar{/}{\kern-.1em}{}}
\g@addto@macro\UrlSpecials{\do : {\Ucolon}}
\makeatother

\begin{document}

\twocolumn[{\csname @twocolumnfalse\endcsname

\maketitle  

\begin{abstract}
\noindent
The detection and classification of intersections between triangles are crucial tasks in a wide range of applications within Computer Graphics and Geometry Processing, including mesh Arrangements, mesh Booleans, and generic mesh processing and fixing tasks. Existing methods are hard-coded and deeply integrated into specific algorithms, and significant efforts are usually required to integrate them into new pipelines or to extend them to different numerical representations. This paper presents a versatile and exhaustive algorithm to identify and classify intersections between triangles with either floating points, rational numbers, or implicit representations. The proposed tool is implemented as a C++ templated and header-only code that is generic and easy to integrate into further algorithms requiring the triangle-triangle intersection detection step. The developed tool has been tested and compared with a state-of-the-art approach, and it is shared with the Geometry Processing community with an Open Source license.

\end{abstract}

\subsection*{Keywords}
Robust Geometry Processing, Triangle-triangle Intersections, Collision Detection, Mesh Arrangements, Exact Arithmetic, Implicit Representations.
\vspace*{1.0\baselineskip}
}]

\copyrightspace

\section{Introduction}
The detection and classification of intersections between triangles represent crucial tasks in several fields of Computer Graphics, Computational Geometry, and Geometry Processing, including mesh Boolean operations, mesh arrangements, mesh repairing, collision detection, 3D reconstruction, and other techniques involving triangle meshes. 

The first complexity of this problem is generated by the significant number of different configurations in which two triangles can intersect each other. If two triangles can intersect into two different points, the complexity becomes particularly challenging in the case of coplanar triangles since we can have up to six different intersection points \rev{(each edge of a triangle crosses two edges of the other triangle)}. Enumerate all the possible configurations is tricky, and missing one of them when implementing a triangle-triangle intersection detection pipeline becomes quite easy.

A second crucial issue emerges when representing points generated by intersections detected in the previous step. Traditional numerical representations based on floating-point arithmetic often lack sufficient precision, causing rounding errors that can propagate throughout the entire algorithm and produce topological inconsistencies in the final result. Indeed, the floating-point space of the representable number leads to the necessity to approximate values, potentially leading to inaccuracies, especially in precision-sensitive computations like Boolean operations on meshes, where even minor errors can propagate into significant issues~\cite{cork2013}.

Existing solutions for intersection detection and classification present significant limitations. First, they are usually deeply integrated with specific algorithms or custom data structures, complicating their reusability in different contexts. Second, most existing solutions are typically limited to a single numerical representation, such as floating-point arithmetic, rational numbers~\cite{cgal}, or implicit points~\cite{Attene_indirect}, restricting the solution's code versatility. A brief overview of numerical representations and their importance in this class of problems will be discussed in Section~\ref{sec:soa}.

To face this problem and facilitate the work of researchers and practitioners in these Computer Graphics areas, we propose an innovative templated tool capable of seamlessly operating across various numerical representations. Our solution, implemented as a header-only C++ library based on templates, works consistently with triangles represented using floating-point arithmetic, rational numbers, or implicit representations. This work represents a practical implementation of a problem discussed in \cite{GSCRedigits}. A central role in our tool is assumed by robust geometric predicates developed by Shewchuk in 1997~\cite{richard1997adaptive}, specifically \textit{Orient2D} and \textit{Orient3D}, which efficiently and exactly determine the relative spatial relationships between a query point and, respectively, a line and a plane. Furthermore, the templated nature of our implementation allows for easy extension to additional numerical representations, allowing our code to be used with any numerical representation implementing the \textit{Orient2D} and \textit{Orient3D} geometric predicates. 

The ease of integration offered by our tool facilitates incorporating the proposed method into new mesh processing tools as well as into existing mesh processing pipelines. To validate the robustness and performance of our solution, we conducted an extensive evaluation process detailed in Section~\ref{sec:validation}. Results show that the performances of our tool are perfectly comparable to state-of-the-art algorithms, with additional advantages in numerical generality and ease of integrability and reusability.

Finally, we released our tool with Open-Source MIT license to encourage and facilitate further research developments in the Computer Graphics community (\href{https://github.com/cg3hci/Fast-and-Robust-Tri-Intersections-with-different-Num-Reprs/tree/main}{Repository \texttt{\textbf{LINK}}}).

\rev{The rest of this paper is structured as follows. In Section~\ref{sec:soa}, we analyze some state-of-the-art works that address the problem of triangle-triangle intersection classification with different numerical systems and different goals. In Section~\ref{sec:tool}, we describe the proposed tool, provide the motivation for the choices made, detail the implementation choices, and explain the use. In Section~\ref{sec:validation}, we describe the testing and validation phase that we performed on our tool, comparing it with the same task performed by a portion of a state-of-the-art algorithm. Finally, in Section~\ref{sec:conclusion}, we describe possible future developments in our work.}

\section{Related works} \label{sec:soa}

Detecting triangle-triangle intersections and representing the generated intersecting points are crucial in several Computer Graphics sub-fields. Typical examples include mesh Arrangements \cite{zhou2016mesh, cherchi2020fast, Guo2024}, a fundamental step for mesh Boolean operations \cite{zhou2016mesh, cherchi2022interactive, Jacobson2013RobustIS, lévy2024exactpredicatesexactconstructions}, or collision detection \cite{ericson2004,Dong2024}, mesh repairing \cite{Attene2013}, 3D reconstruction \cite{Trettner2022}, and several other tasks. Even minor inaccuracies in intersection detection or point representation can lead to significant topological problems, damaging the correctness and robustness of the entire algorithm pipeline.

Despite the variety of proposed algorithms in literature and reference implementations on the web, a recurring issue is the lack of a generic solution that works with different numerical representations and that can be easily integrated into newly developed algorithms. Indeed, existing methods cannot be easily adapted or reused across different scenarios without substantial modifications. Adaptation generally requires aligning the intersection detection strategy with different numerical representations or different data structures.

Several applications that detect intersections between triangles can (sometimes) sacrifice precision in favor of the execution time, as we often have in collision detection pipelines. However, there are pipelines, such as mesh Arrangements, in which precision and robustness are fundamental since even minor approximation errors can lead to completely wrong results.

Mesh Arrangements involve partitioning the space defined by a set of triangles into a valid simplicial complex, necessitating robust intersection detection and classification to avoid degeneracies or overlaps. In general, in a mesh Arrangements pipeline, we need two fundamental steps: (1) locate all the intersecting elements and (2) split them in order to create a new structure (via re-triangulation) which includes intersection points and segments~\cite{LCSA_linear_earcut}. Inaccurate handling of intersections can compromise the entire mesh structure, making the mesh unusable for subsequent applications. Both the intersection detection and the creation of the new topology require absolute precision, which can be achieved with different technologies. Zhou et al. \cite{zhou2016mesh} leverage adaptive predicates to detect the intersecting elements, then use rational numbers to ensure absolute precision in the representation of the coordinates of the new points. Cherchi et al. \cite{cherchi2020fast} improved upon this approach using geometric predicates~\cite{richard1997adaptive} to detect the intersections, and then they rely on an implicit representation of the intersection points~\cite{Attene_indirect} gaining execution speed without losing robustness. Mesh Boolean operations similarly require exact intersection detection to accurately determine internal and external model regions. Errors in intersection detection may lead to ambiguous or incorrect outcomes. \rev{All the above works highlight the need to develop algorithms that are robust to numerical approximation errors by resorting to different types of exact representations.}

The first issue in an intersection detection pipeline is ensuring absolute robustness while maintaining an efficient algorithm in terms of time and memory. Among the first fast algorithms for triangle-triangle intersection detection, Möller \cite{moller1997fast} and Held \cite{Held01011997} proposed two algorithms, both based on checking intersections between a line and the planes containing the triangles. Although very fast, these algorithms are not particularly robust in precision, but they are a fundamental groundwork for future methods. Guigue and Devillers achieved significant advancement in robustness \cite{guigue2003fast}, who introduced an algorithm relying solely on the signs of geometric predicates. This method avoids constructing intersection lines or segments between planes, thus reducing potential sources of error. Ling-Yu Wei \cite{Ling-yu2014} introduced further optimizations to reduce the average computational cost by improving the management of non-intersecting cases and using vectorized operations, making the method suitable for real-time applications in animation and virtual reality. Intersection detection algorithms remain a thoroughly explored topic in literature. We refer the reader to \cite{skala_2023} for a comprehensive overview of this class of algorithms.

Fundamental ingredients in this kind of algorithm are (1) the detection of the intersections and (2) the computation of the new points generated by the intersecting elements.

\subsection*{Geometric Predicates}
Fundamental tools of most of the algorithms that detect the intersection between triangles are the Shewchuk geometric predicates \cite{richard1997adaptive} introduced in 1997. Among them, two are particularly interesting for this class of problems. The first, called $Orient2D$, takes as input two points describing a straight line and a third one as a query point. Thanks to the evaluation of the sign of a determinant, this predicate is able to evaluate whether a point is located in one of the two half-planes ($Orient2D<0$ or $Orient2D>0$) or exactly on the line ($Orient2D=0$). Analogously, the $Orient3D$ geometric predicate takes as input three points defining a plane and a fourth one as a query point, and it determines whether the point is located in one of the two half-spaces ($Orient3D<0$ or $Orient3D>0$) or precisely on the plane defined by the three points ($Orient3D=0$). Combining these predicates makes it possible to quickly and robustly identify (without numerical precision problems) the intersections between the simplexes representing the two tested triangles. Shewchuk's predicates operate using an adaptive arithmetic technique. Initially, the expression is evaluated using standard floating-point operations. If the computed value exceeds a predefined safety threshold, the result is returned immediately, ensuring efficiency. In cases where the outcome falls near the uncertainty threshold, the method progressively increases the arithmetic precision, guaranteeing an exact result without resorting to full multiprecision arithmetic for every operation.

It is worth noting that these algorithms are not limited to intersection detection; they can also be used for classification. For instance, if an algorithm detects that triangles intersect, it can typically identify the edges involved in the intersection (calculating intersection points or segments). In the case of non-coplanar triangles, the intersection, if it exists, can only be a point or a linear segment; however, coplanar triangles can overlap in polygonal areas. Nevertheless, managing edge cases (e.g., triangles just touching at a vertex or lying on the same plane with overlapping edges) is delicate. Algorithms based on standard floating-point arithmetic may yield inconsistent results (e.g., a triangle "barely touches" another but, due to numerical errors, either appears entirely separate or with slight penetration). Algorithms based on geometric predicates provide an excellent foundation for robustness by reducing every decision to a determinant sign comparison. 

\begin{figure}
\centering
\includegraphics[width=1\linewidth]{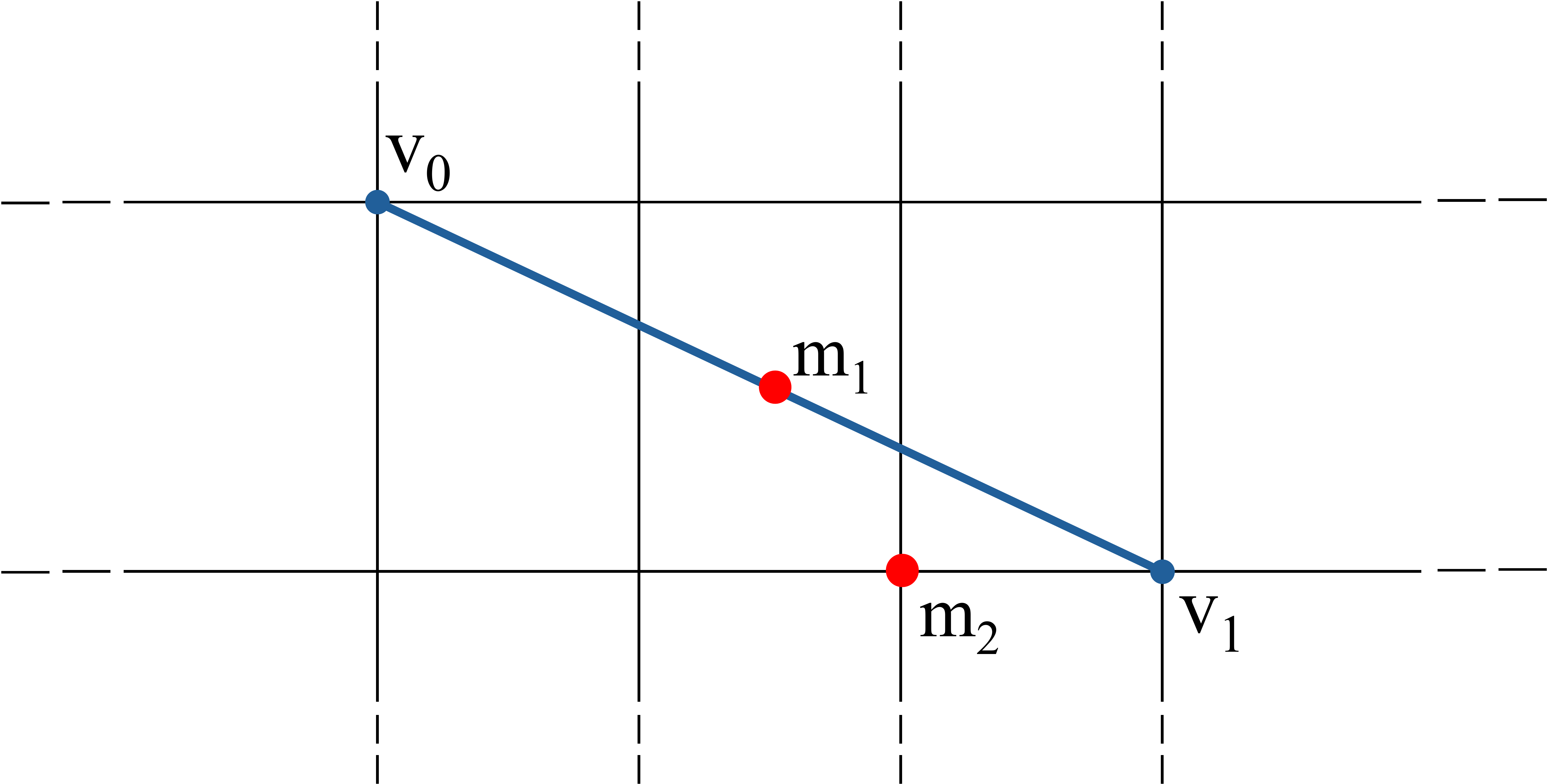}
\caption{An example of approximation error with the floating-point notation. We can see the space of possible values representable by floating-point numbers as a grid. Even a simple operation like computing the midpoint of a segment can produce a point that cannot be precisely represented in the floating-point grid. This example shows that the exact midpoint \rev{$m_1$} of the segment $V_0 - V_1$ is approximated to the closest representable point \rev{$m_2$.}}
\label{fig:fp_error}
\end{figure}

As we stated before, locating and classifying the intersection is not the only problematic step of this class of algorithms. After finding the intersecting elements, we need to represent the generated intersection points to use them in the remaining flow of the algorithm. The most common method employs floating-point arithmetic, approximating the coordinates of the new point into the closest representable number. Indeed, by construction, the space of numbers we can represent with floating-point notation can be seen as a non-uniform grid with an increased density in proximity to zero. Even a simple case, e.g., calculating the midpoint of a segment (see Figure~\ref{fig:fp_error}), might yield a midpoint approximated to the nearest representable value, potentially causing errors propagating through subsequent pipeline steps. While floating-point numbers provide speed in computation, they cannot guarantee precision in numerical representation. For this reason, sometimes, we decide to renounce performance in favor of absolute precision.

\subsection*{Alternative Numeric Representation}

A famous and spread alternative method to represent the coordinates of the new points is to use Rational numbers~\cite{zhou2016mesh}. Rational numbers encode values as fractions with arbitrary-length numerators and denominators, providing absolute precision~\cite{cgal}. However, rational arithmetic's computational complexity and performance overhead often limit their practical use in performance-critical applications. Consider that even a simple operation like a sum between two values (represented as a potentially huge fraction) can become highly resource-demanding. 

In some cases, the explicit coordinates of the new points are not necessary to continue the algorithm flow. When we detect the intersection between the triangles simplexes, we can decide not to explicitly represent the intersection point's coordinates in order to avoid approximation errors. In this case, we represent intersection points explicitly by storing the coordinates of the original input elements involved in the intersection. For example, the intersection of a segment (edge of a triangle) and a triangle can be stored as "\textit{LPI}" point, i.e. , a "Line-Plane Intersection". In this case, we store the coordinates of the two segments' endpoints and of the three triangles' corners rather than the coordinates of the points generated by their intersection. Representing intersection points in this indirect way also requires re-designing the geometric predicates, which has led to the development of "Indirect" predicates \cite{Attene_indirect}. This new family of geometric predicates can compute the same information as the original Shewchuk's geometric predicates but works on implicitly represented points. This approach maintains robustness against rounding errors and offers a practical balance between computational efficiency and the stringent accuracy demanded by many advanced mesh processing tasks.

Countless algorithms in the literature deal with finding the intersections between triangles. Considering that there are different ways to represent the intersection points and, consequently, different versions of the geometric predicates that operate with them, we feel that a tool like the one we propose could be extremely useful for the scientific community in these fields. In fact, our tool allows for easy integration into new algorithms under development and supports multiple numerical representations. Furthermore, by design, our tool allows for easy extension with additional numerical representations.

\section{the tool}
\label{sec:tool}

\begin{figure}
    \centering
    \begin{subfigure}[b]{0.32\linewidth}
        \centering
        \includegraphics[width=\linewidth]{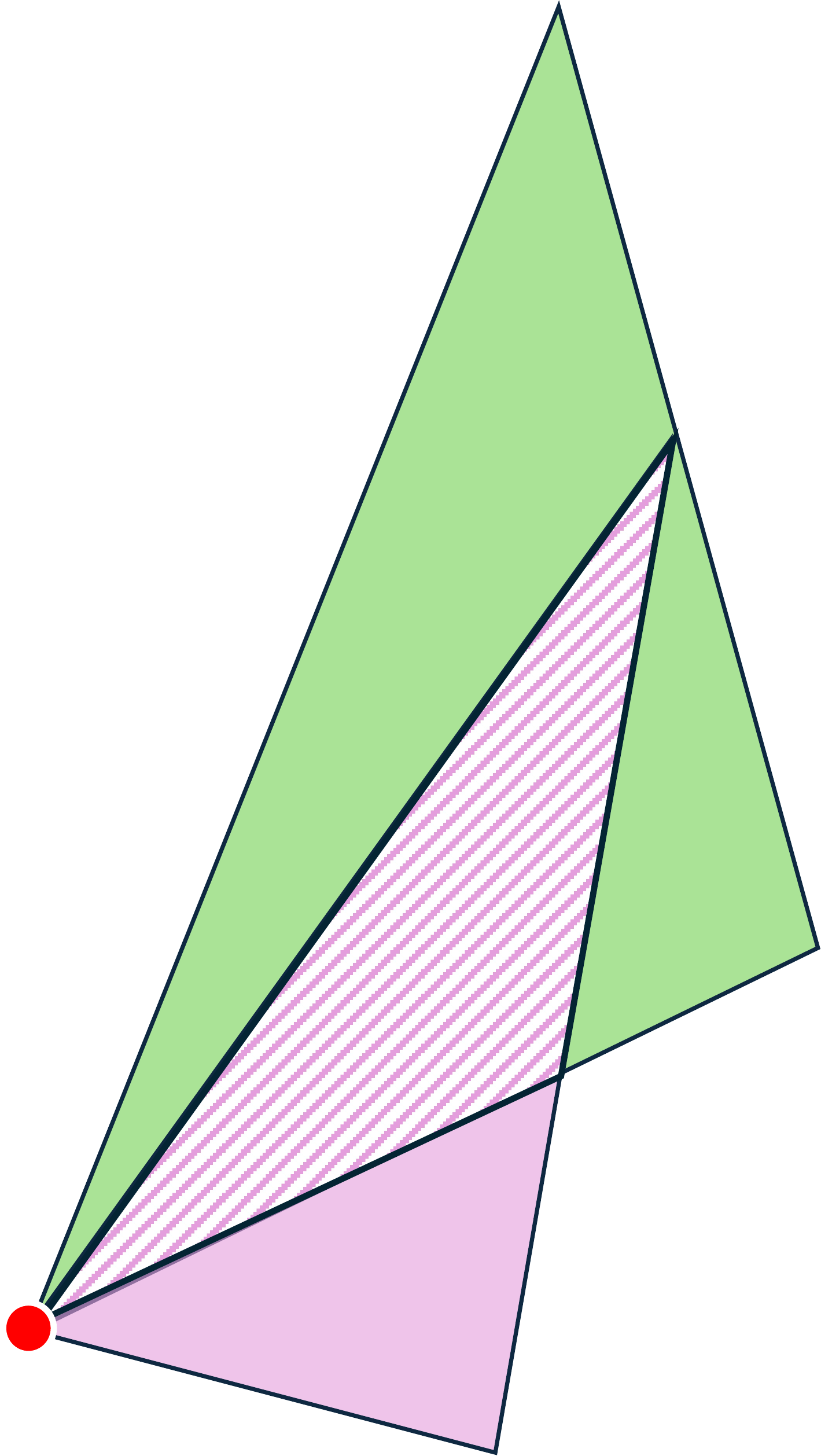}
        \caption{}
        \label{subfig:t1}
    \end{subfigure}
    \hfill
    \begin{subfigure}[b]{0.32\linewidth}
        \centering
        \includegraphics[width=\linewidth]{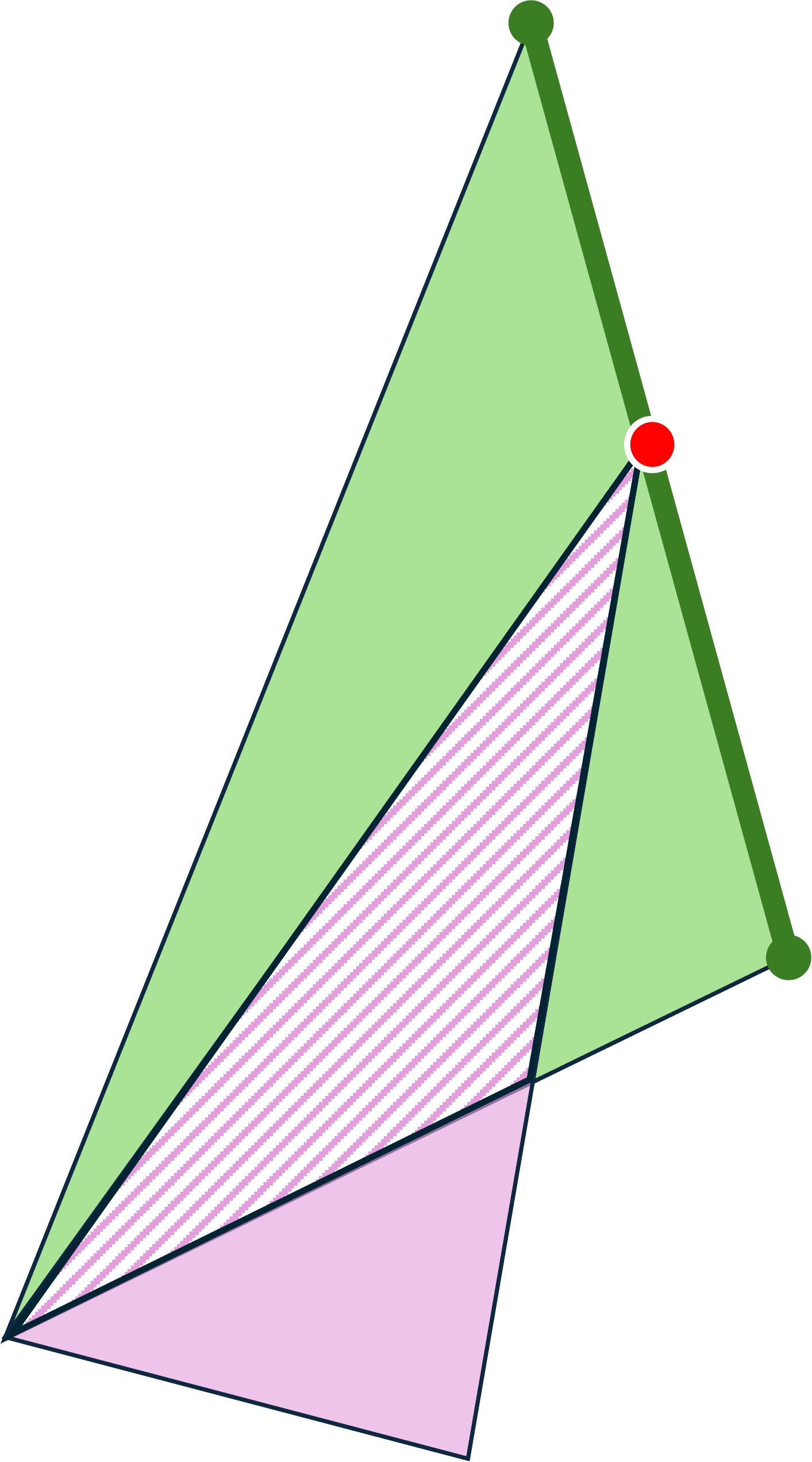}
        \caption{}
        \label{subfig:t2}
    \end{subfigure}
    \hfill
    \begin{subfigure}[b]{0.32\linewidth}
        \centering
        \includegraphics[width=\linewidth]{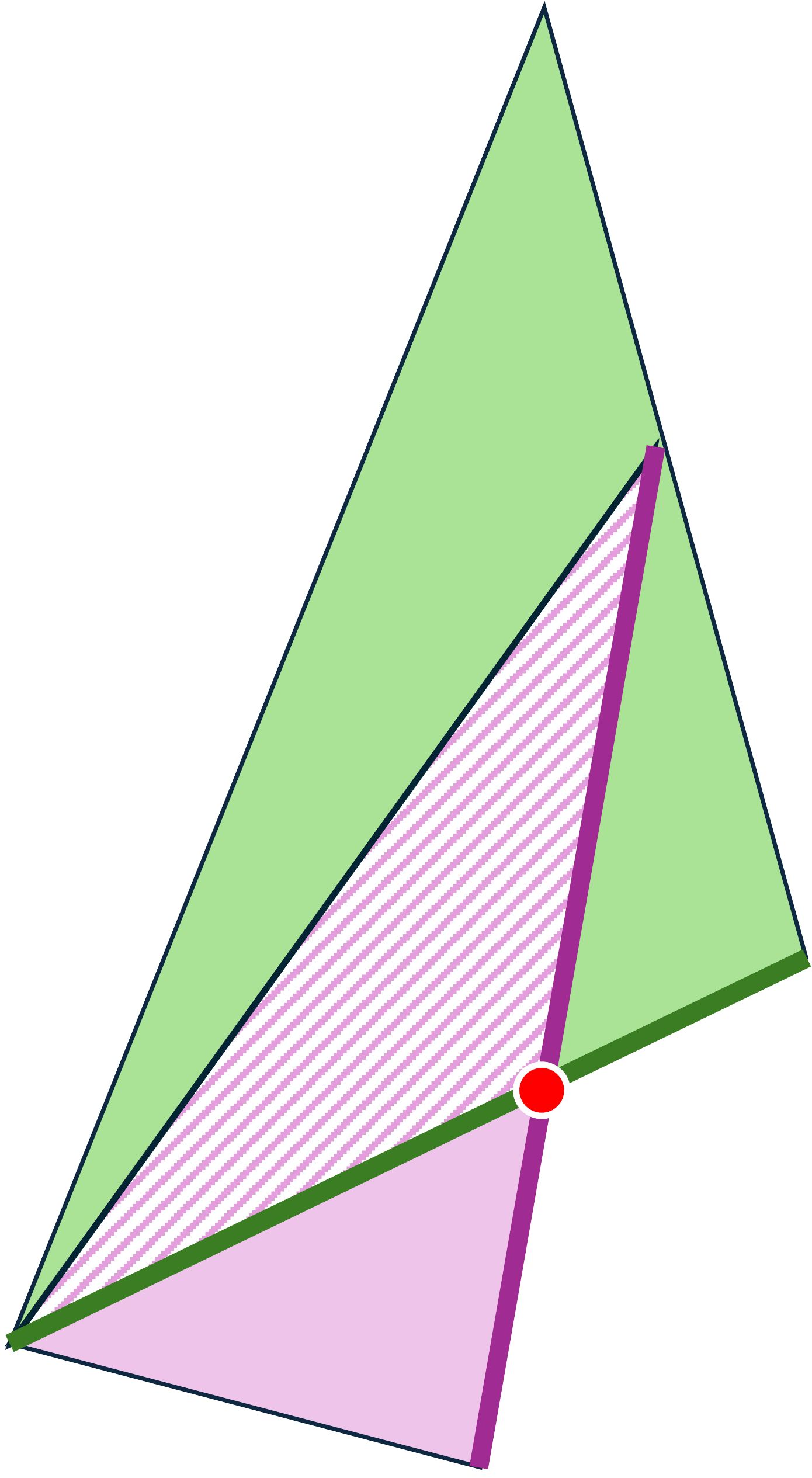}
        \caption{}
        \label{subfig:t3}
    \end{subfigure}
    \\[0.5cm]
    \caption{Decomposition of an intersection between two triangles into intersections between simplexes of smaller dimension: (a) coincident points, (b) point in segment, and (c) segment crossing segment.}
    \label{fig:coplanar_intersec}
\end{figure}

We present an easy-to-use tool written in C++ as a header-only library. It takes as input the coordinates of two triangles and returns information about the identified intersections. The core of the tool is divided into two main tasks: (1) detect and classify the intersections, and (2) return the information to create the new intersection points with a chosen numerical representation. To address the first step, we split the problem into simplexes of lower dimensions, meaning that the intersections between two triangles are identified by checking the intersections between faces, segments, and points~\cite{lévy2024exactpredicatesexactconstructions}. An example of this process can be seen in Figure~\ref{fig:coplanar_intersec}, when a first intersection is defined by two coincident vertices, a second one is defined by a point lying on a segment, and a third one is generated by a segment of the first triangle crossing a segment of the second one.

To locate all possible intersections between the simplexes of the triangles in a robust and fast way, we use the \textit{Orient2D} and \textit{Orient3D} predicates~\cite{richard1997adaptive} included in the \cite{cinolib} library. Indeed, thanks to the combination of these two predicates, we can exactly understand if two simplexes touch or intersect each other. Since we aim to allow this identification process in triangles represented with different numerical representations, we developed our tool with C++ templates in order to support every kind of representation that is able to implement the \textit{Orient2D} and \textit{Orient3D} geometric predicates. In our current implementation, we already provide the support for the classical floating-point numbers with the original Shewchuck predicates~\cite{richard1997adaptive}, the rational numbers implemented in the CGAL library~\cite{cgal}, and the implicit points introduced in \cite{Attene_indirect}.

\begin{figure}[h]
    \centering

    \begin{subfigure}[t]{0.3\linewidth}
        \centering
        \includegraphics[width=\linewidth]{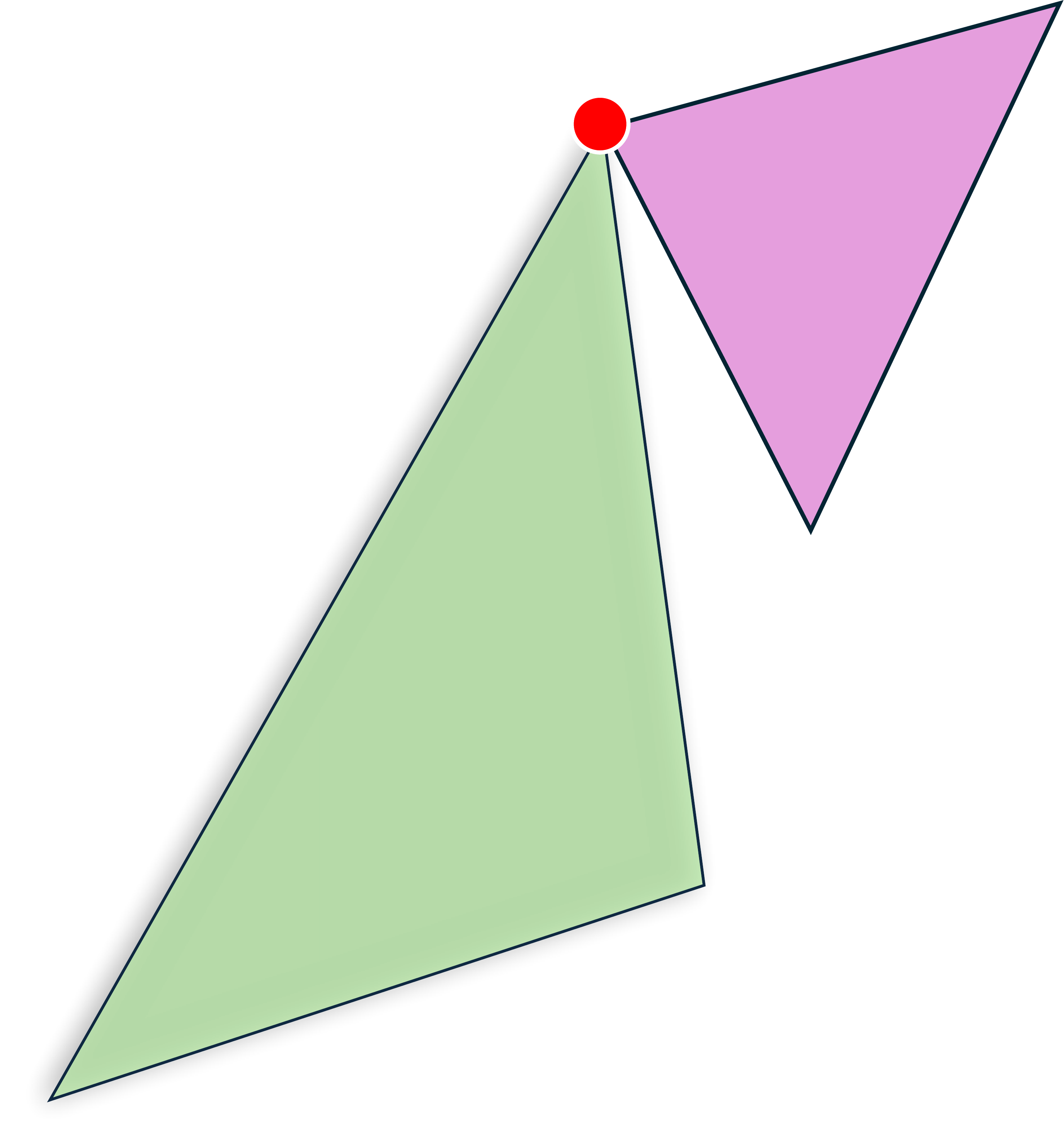}
        \caption*{\footnotesize{Coincident Points}}
    \end{subfigure}
    \hfill
    \begin{subfigure}[t]{0.3\linewidth}
        \centering
        \includegraphics[width=\linewidth]{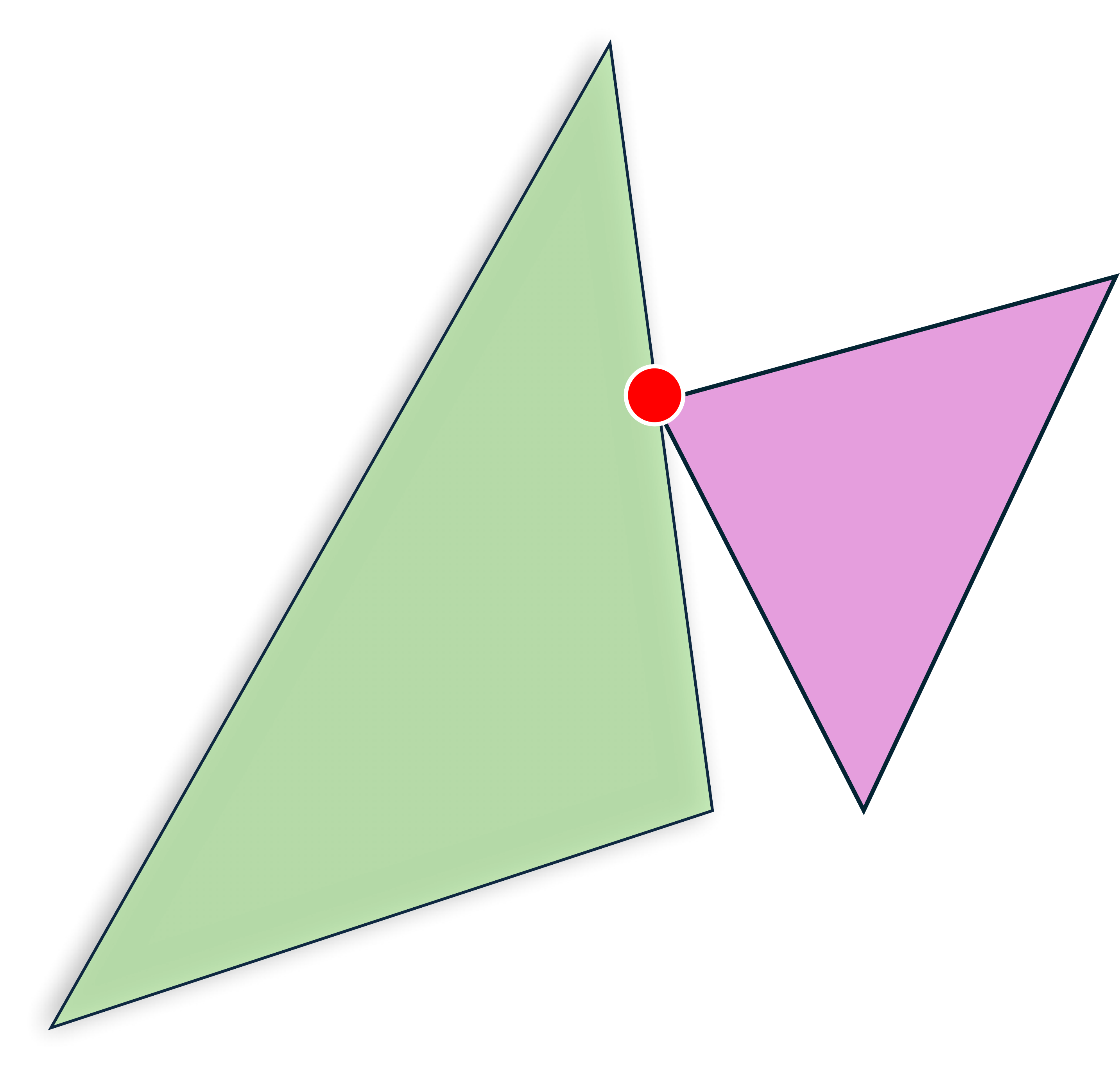}
        \caption*{\footnotesize{Point in Segment}}
    \end{subfigure}
    \hfill
    \begin{subfigure}[t]{0.3\linewidth}
        \centering
        \includegraphics[width=\linewidth]{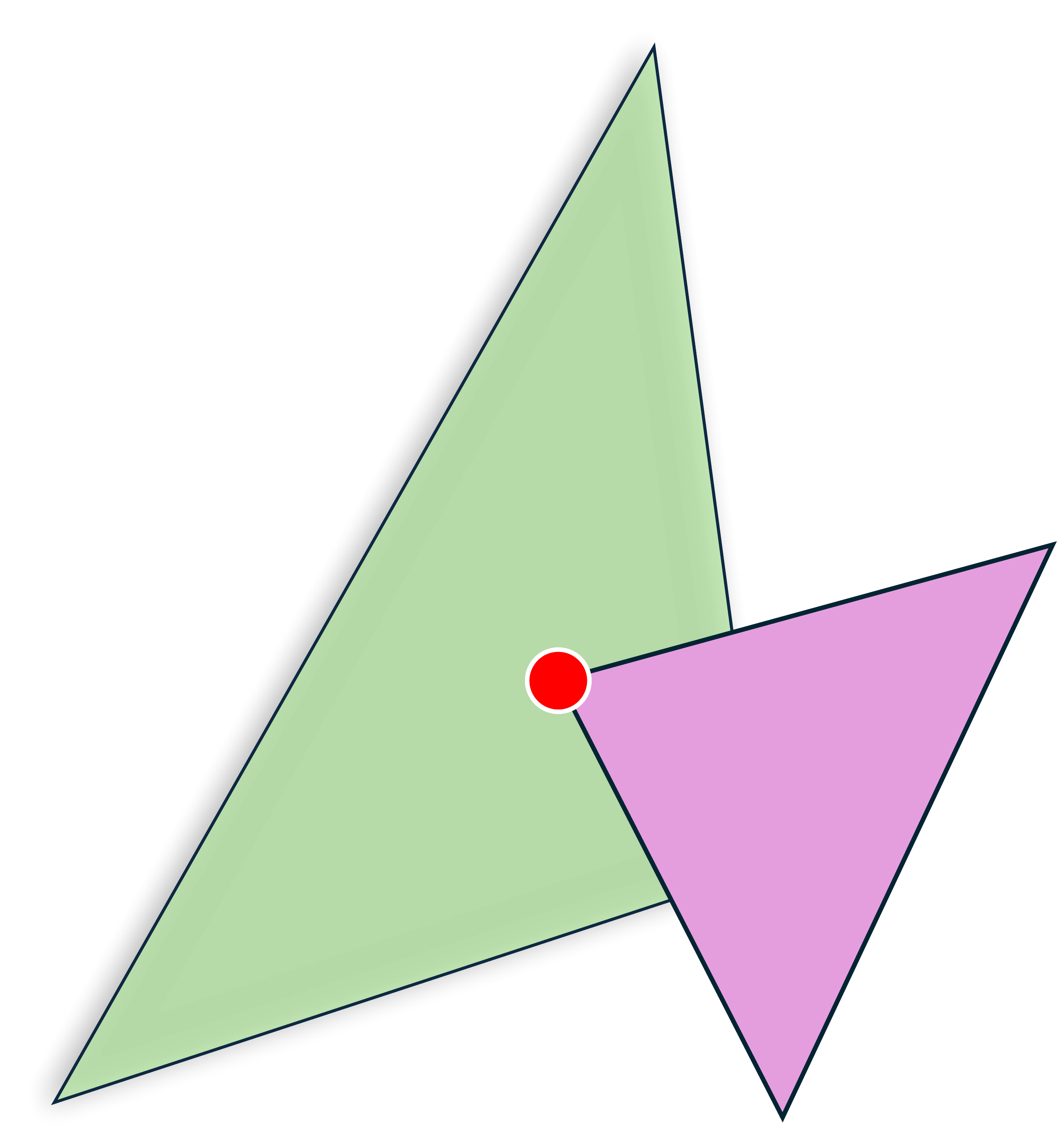}
        \caption*{\footnotesize{Point in Triangle}}
    \end{subfigure}

    \vspace{0.5em} 

    \begin{subfigure}[t]{0.45\linewidth}
        \centering
        \includegraphics[width=.7\linewidth]{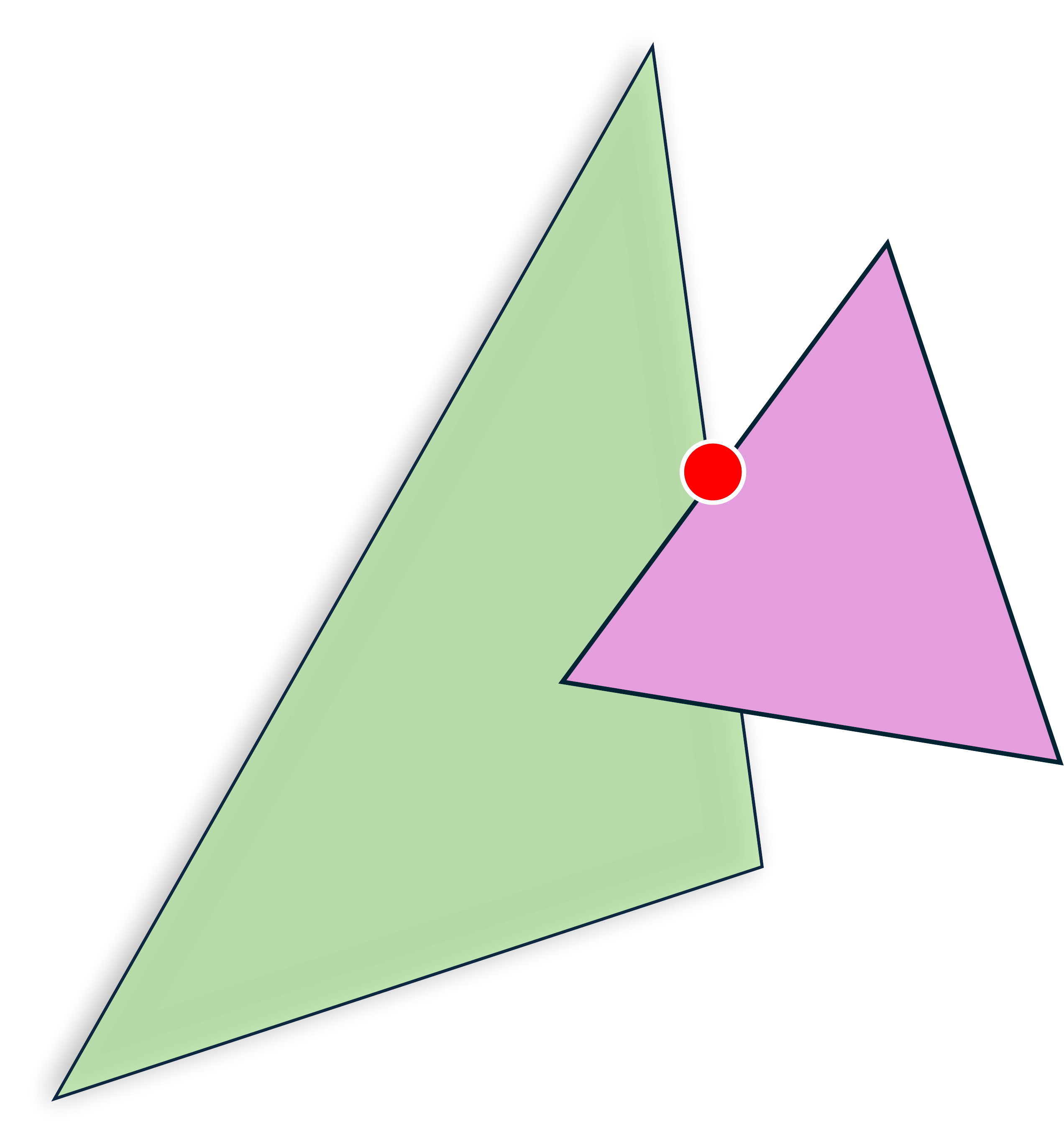}
        \caption*{\footnotesize{Segment crossing Segment}}
    \end{subfigure}
    \hspace{0.05\linewidth}
    \begin{subfigure}[t]{0.45\linewidth}
        \centering
        \includegraphics[width=.7\linewidth]{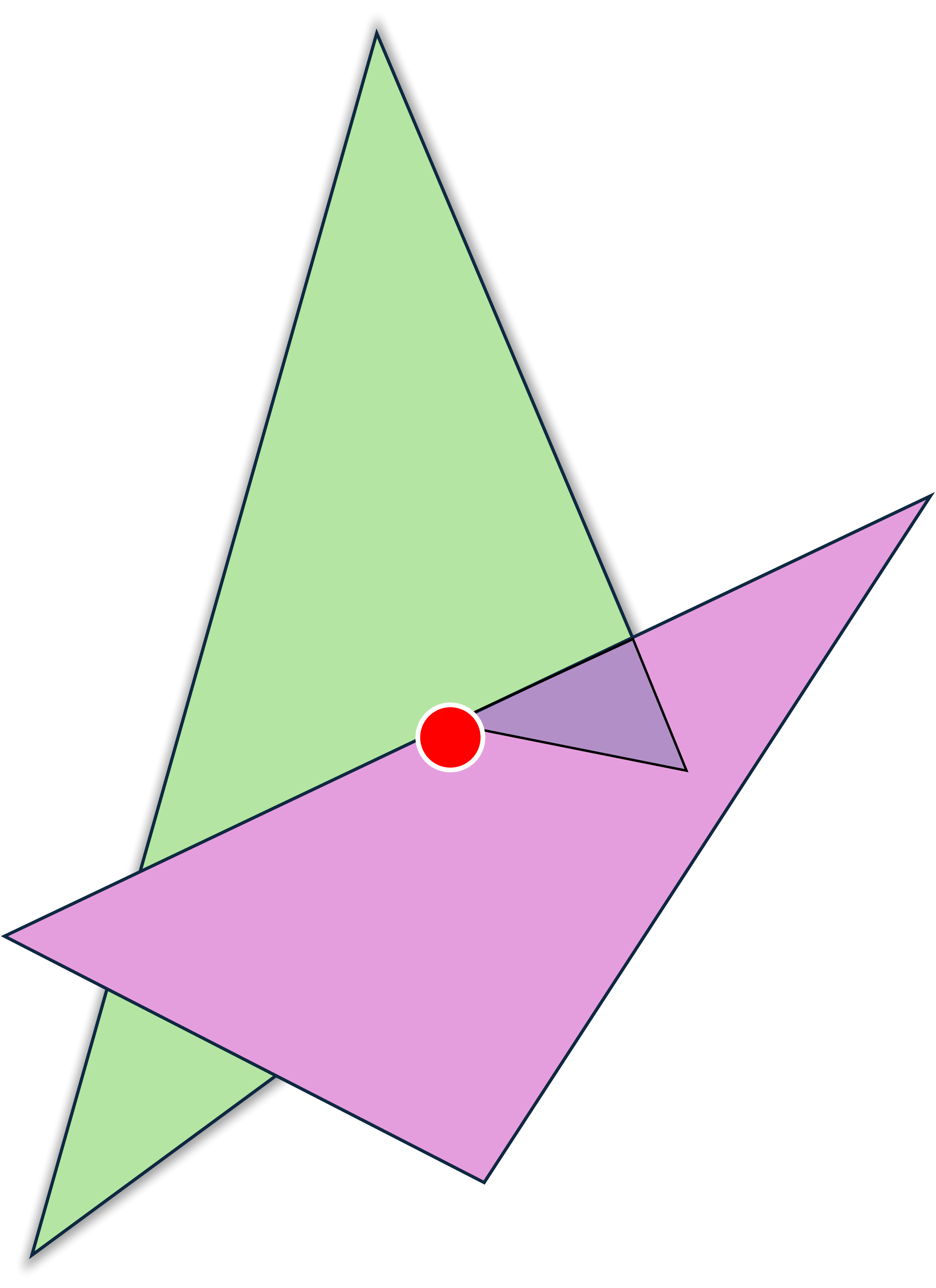}
        \caption*{\footnotesize{Segment crossing Triangle}}
    \end{subfigure}
    \vspace{0.5em}

    \caption{\rev{All possible cases of intersections between two triangle sub-simplexes.}}
    \label{fig:five_intersections}
\end{figure}

We classify all possible intersections into a set of intersecting simplexes, which we also return as output information. In particular, we can have (1) \textit{coincident points}, (2) \textit{point in segment}, (3) \textit{point in triangle}, (4) \textit{segment crossing segment} and (5) \textit{segment crossing triangle} \rev{(see Figure \ref{fig:five_intersections} for a mosaic of all the five described cases)} .

\begin{figure}[h]
    \centering
    \includegraphics[width=1\linewidth]{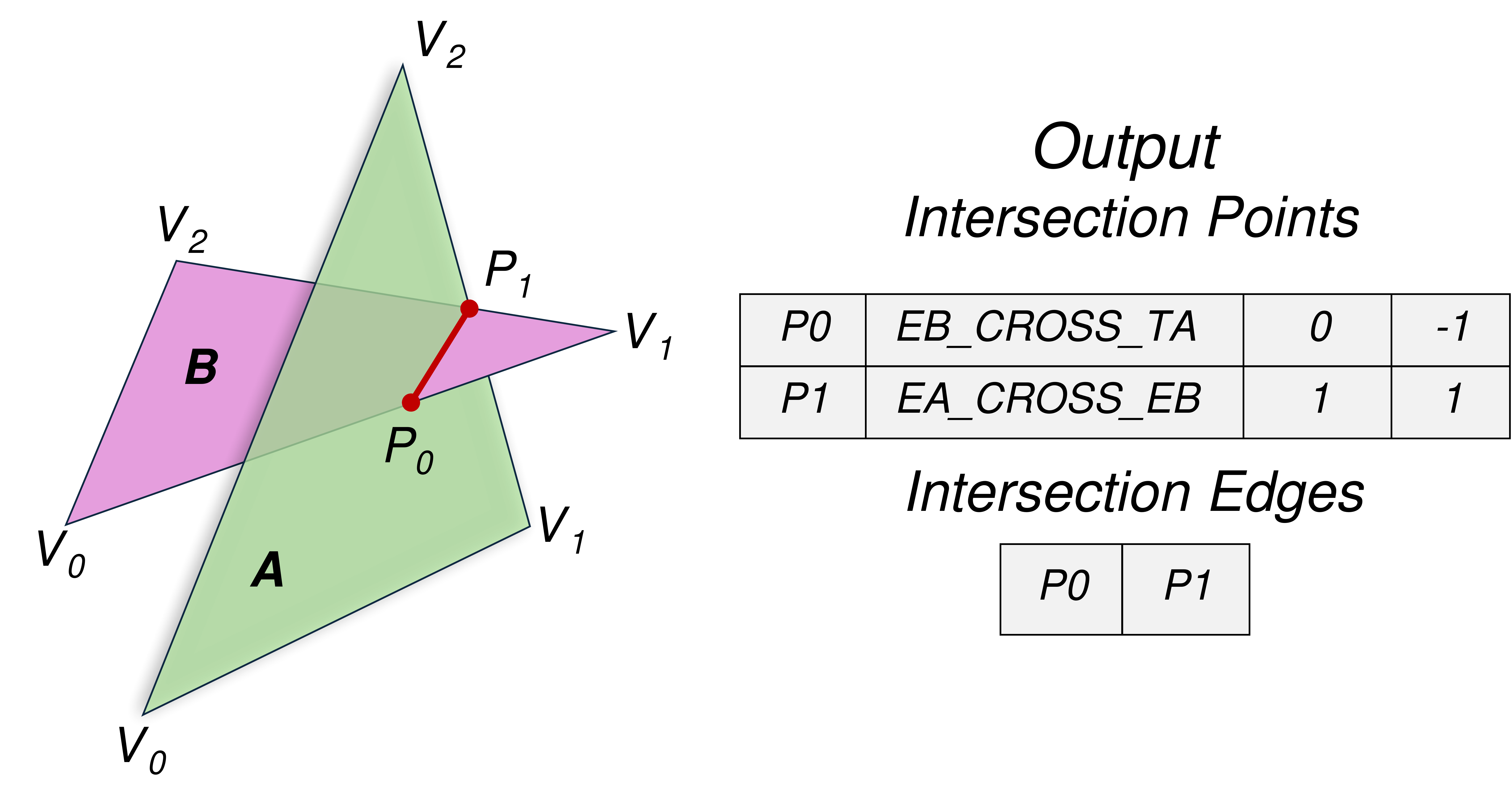}

    \caption{An example of an intersection between two triangles and the corresponding output representation. \rev{Edge $V_0 - V_1$ of triangle B crosses triangle A generating the $P_0$ intersection point. Point $P_1$ is given by the intersection between edge $V_1 - V_2$ of triangle A and edge $V_1 - V_2$ of B.} Points $P_0$ and $P_1$ form an intersection edge. Note that, if less than two ids are required to express the intersection, we fill the useless field with -1.}
    \label{fig:tri_tri_output}
\end{figure}

As a starting point of the detection pipeline, we catch, with six \textit{Orient3D} calls, whether the input triangles are coplanar or not, or if they do not intersect. In the case of non-coplanar triangles, we can find up to two intersections among the ones listed before, while in the case of coplanar triangles, we can find up to six intersections, excluding case (5).

\begin{figure}[h]
    \centering
    \begin{subfigure}[]{0.48\linewidth}
        \centering
        \includegraphics[width=\linewidth]{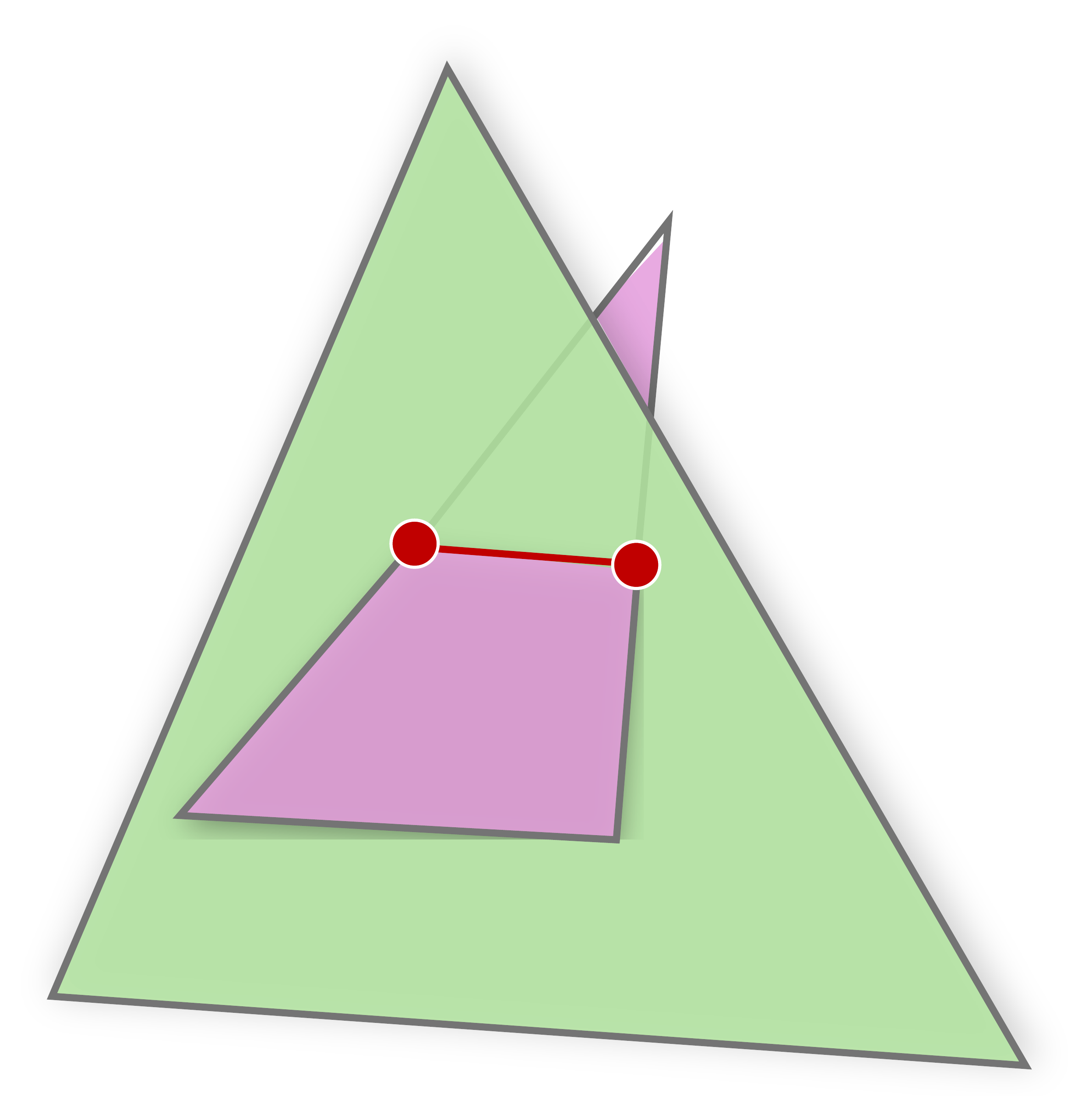}
        \caption{}
        \label{subfig:noncoplanar}
    \end{subfigure}
    \hfill
    \begin{subfigure}[]{0.48\linewidth}
        \centering
        \includegraphics[width=\linewidth]{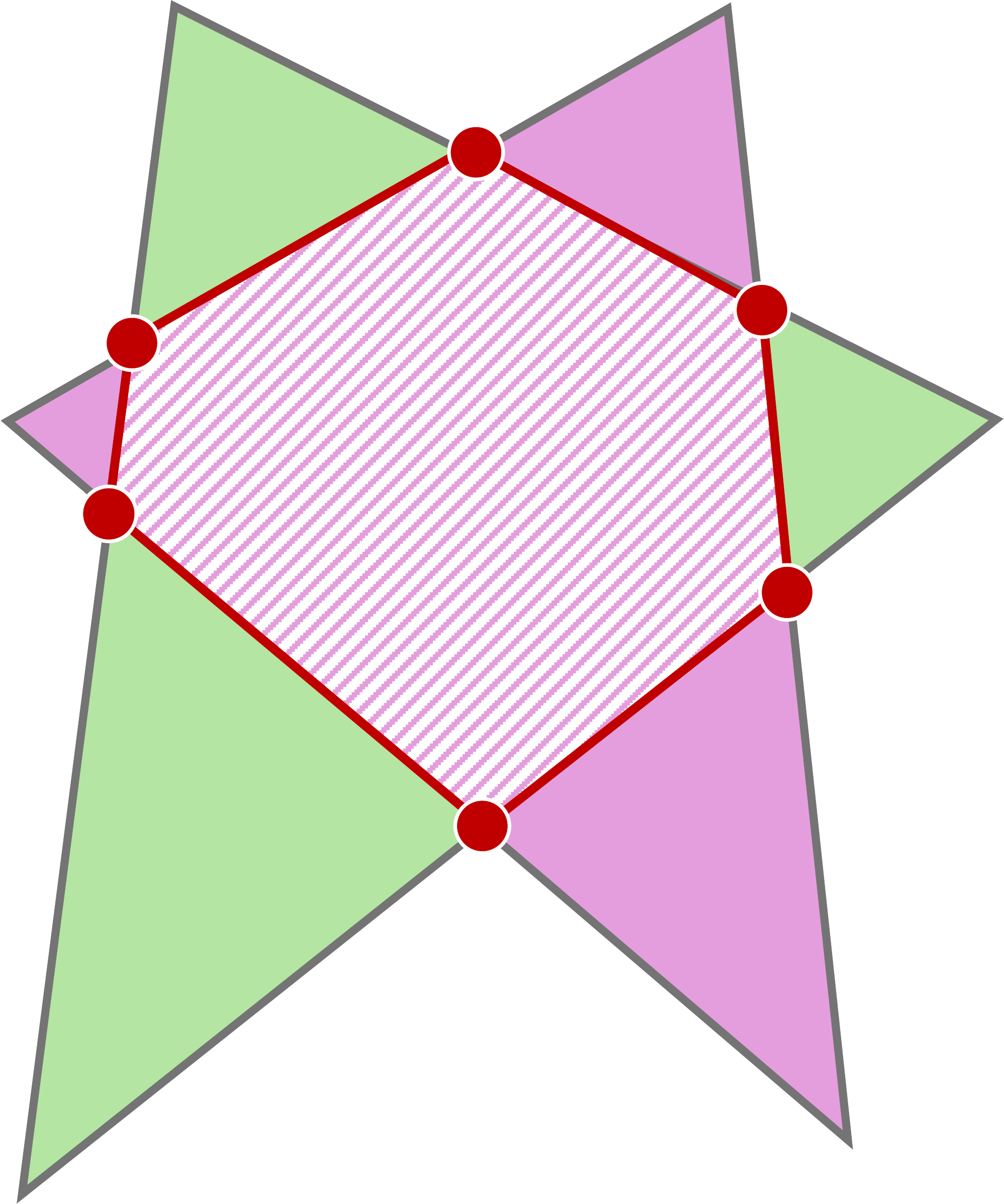}
        \caption{}
        \label{subfig:coplanar}
    \end{subfigure}
    \\[0.3cm] 
    \caption{Examples of intersecting triangles. In (a), two non-coplanar triangles generate two intersection points connected by an intersection segment. In (b), two coplanar triangles produce six intersection points connected pairwise into segments, forming a planar polygon.}
    \label{fig:segments}
\end{figure}

Furthermore, intersections between two triangles can generate not only intersection points but also "\textit{intersection segments}". Indeed, if you focus on Figure~\ref{fig:tri_tri_output}, the intersection between triangles A and B generated two intersection points linked together to form a new segment. For most of the algorithms requiring the classification of the intersections between triangles, intersection segments are also important information. For this reason, we also collect this information and return it as output together with the intersection points. Note that, finding the intersection segments can be a tricky problem. In fact, in the simplest case of non-coplanar triangles, we can have up to two intersection points and only a single intersection segment connecting them (Figure~\ref{subfig:noncoplanar}). However, in the more complex case of coplanar triangles, we can combine up to six intersection points into intersection segments in several ways, leading to a planar polygon composed of up to six sides (Figure~\ref{subfig:coplanar}).

After finding the intersection points and segments, we are ready for the second step of the algorithm: returning all the information about the found intersections. Each triangle has been divided into lower simplexes, so we package the intersections as pairs of simplexes that generated it, together with a label providing information about the intersection type. In this way, new intersection points can be created with the numerical representation we select, whether explicit (e.g., floating-point or rational numbers) or implicit (implicit points).  

\begin{algorithm}[]
\begin{algorithmic}[1]
\Require List of vertices $V$ of two triangles $T_0$ and $T_1$
\Ensure List $L$ containing information about the detected intersection points and segments 
\\
\State Compute orientations of $T_0$ vertices w.r.t. $T_1$ plane
\State Compute orientations of $T_1$ vertices w.r.t. $T_0$ plane

\If{\rev{all vertices of a triangle are on the same side of the other triangle plane}}
\State \Return $L$ (empty, no intersections)
\EndIf\\


\ForAll{vertices \rev{$v_i$} in $V$}
    \State check coincident vertices    
    \State add coincident vertices in $L$
    \\

    \If{$v_i$ lies on an edge or area of the other triangle}
    \State add $v_i$ in $L$
    \EndIf
\EndFor\\

\ForAll{ edges $e_i$ of $T_0$ and \textbf{all} edges $e_j$ of $T_1$}
    \If{$e_i$ crosses $e_j$}
        \State add info about intersection in $L$
    \EndIf
\EndFor\\

\If{$T_0$ and $T_1$ are not coplanar}
    \ForAll{edge $e_i$ of $T_0$ or $T_1$}
    \If{$e_i$ crosses the area of other triangle}
        \State add info about intersection in $L$
        \EndIf
    \EndFor\\
    \If{$L$ contains 2 intersection points}
        \State add intersection segment $(p_0,p_1)$ in $L$
        \EndIf
\EndIf\\

\If{$T_0$ and $T_1$ are coplanar}
    \ForAll{point combinations ($p_i$, $p_j$) in $L$}
        \If{$p_i$ and $p_j$ lies on the same input edge of $T_0$ or $T_1$}
            \State add intersection segment $(p_i,p_j)$ in $L$
        \EndIf
    \EndFor

\EndIf\\

\State \Return $L$
\end{algorithmic}
\caption{\textbf{Triangle-Triangle Intersection}. The algorithm identifies intersections by checking vertices, edges, and areas intersections. It takes as input the six vertices of two triangles $T_0$ and $T_1$ and returns a list $L$ containing all the information about the detected intersections.}
\label{alg:tri-tri-detection}
\end{algorithm}

The output structure of our tool is composed as follows:

\begin{itemize}
    \item A list of intersection points, each one represented as a tuple in the form \texttt{(type, id0, id1)}, where \texttt{type} indicates the kind of intersection and the indices \texttt{id0} and \texttt{id1} to identify the simplexes involved in the intersection.
    
    \item A list of intersection segments, represented as tuples \texttt{(p0, p1)}, which refer to the positions of the intersection points inside the previously described list.
    
    \item Additional metadata providing supplementary information, such as whether the investigated triangles are coplanar, or other customizable information.
\end{itemize}

In Figure~\ref{fig:tri_tri_output}, we can see an example of our output structure, which describes the intersection of two non-coplanar triangles. In Algorithm~\ref{alg:tri-tri-detection}, we sum up all the steps of the described tool in pseudo-code.

\section{validation} \label{sec:validation}

We tested our tool with an empirical validation involving the intersection detection on thousands of triangle meshes from the famous Thingi10k dataset~\cite{Thingi10K}, which is widely recognized in the literature for its high complexity \cite{lazard:hal-04907149}.

To evaluate the robustness and efficiency of the tool, we included our code in the public reference implementation of a state-of-the-art mesh Arrangement pipeline~\cite{cherchi2020fast}, replacing the original detection and classification module (i.e., the function \texttt{checkTriangleTriangleIntersections} inside the  \texttt{classifyIntersections} function in the public code). 

In the experimental setup, conducted \rev{workstation equipped with 12 cores intel I9 processor and 128GB of RAM}, we compared the execution time of our detection algorithm with that of the original pipeline, imposing a maximum runtime of 15 minutes per model and collecting data about 9996 models. 

We checked if our tool produces precisely the same intersection list as the state-of-the-art algorithm and if the final result is exactly the same. To do so, we listed all the obtained intersections (both points and segments) in a common format, and we compared the resulting arranged mesh in terms of the number of simplexes and Euler's characteristic.  

\begin{figure}[h]
    \centering
    \includegraphics[width=1\linewidth]{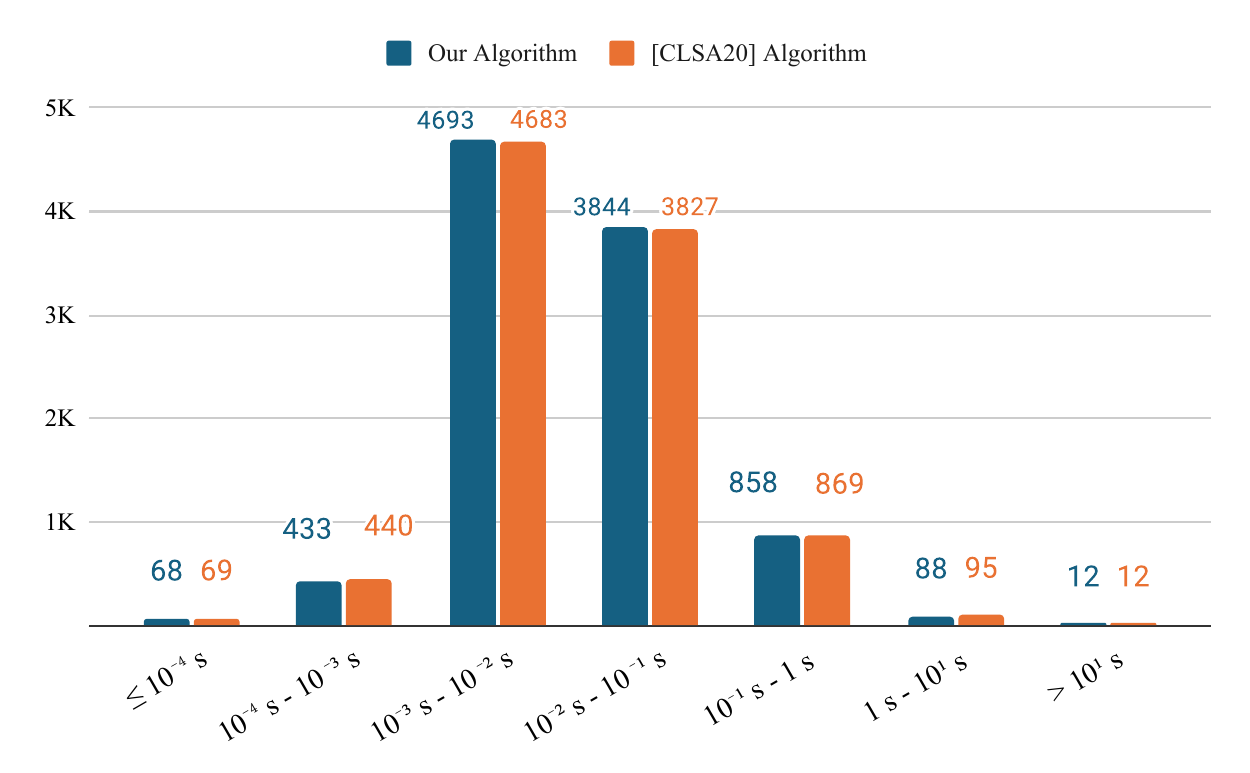}
    \caption{In this bar chart, we can see the models taken from Thingi10K dataset~\cite{Thingi10K} processed by our algorithm and the one of \cite{cherchi2020fast}, split into classes based on the time required for the detection and classification steps. On the horizontal axis we report the execution time intervals of the intersection detection and classification module, while on the vertical axis we report the number processed models. Blue bars (ours) and orange bars (\cite{cherchi2020fast})represent the number of models processed in that time interval with the two tested algorithms. As you can easily see, the number of models processed for each time interval is the same.}
    \label{chart:grafico_1}
\end{figure}

\begin{figure}[h]
    \centering
    \includegraphics[width=1\linewidth]{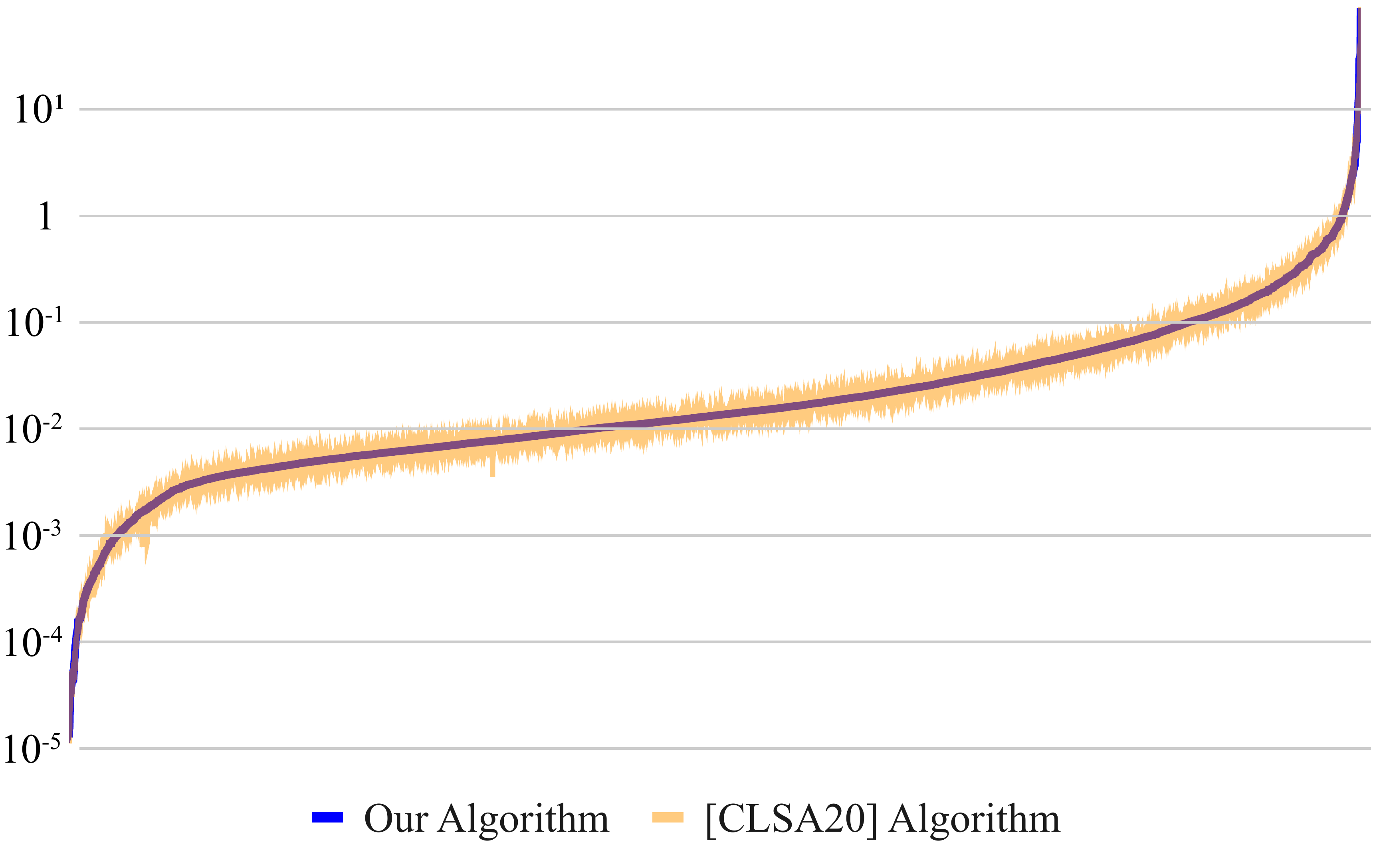}
     \caption{In this line chart, we directly compare the execution times of our detection and classification pipeline w.r.t. those of \cite{cherchi2020fast} on the models of Thingi10K dataset~\cite{Thingi10K}, in logarithmic scale. In the horizontal axis, \rev{we point out that the models are sorted by increasing time with respect to our algorithm}, which is indicated in the vertical axis. Also in this case, the execution times of the two algorithms are almost identical.}
    \label{chart:grafico_2}
\end{figure}

Once we were sure that the result obtained was correct, we compared the execution time of the two algorithms. As can be seen from the  charts in Figure~\ref{chart:grafico_1} and Figure~\ref{chart:grafico_2}, the two algorithms are fully comparable in terms of execution times.

In Figure~\ref{chart:grafico_1}, we show the models taken from Thingi10K dataset~\cite{Thingi10K} processed by our algorithm and the state-of-the-art comparison~\cite{cherchi2020fast}, split into classes based on the time required for the detection and classification steps. Each bar in the chart represents the number of models processed in that time interval. As is clearly visible, the number of models processed for each analyzed time interval is almost the same.

The same behavior is shown in the chart in Figure~\ref{chart:grafico_2}, in which we report a line chart showing a direct comparison of the execution times of the two algorithms. We used the logarithmic scale to emphasize the timing. The models are sorted by increasing time (of detection and classification). Also in this case, we can say that, except for almost imperceptible oscillations, the execution times of the two algorithms are almost identical.

\begin{table*}[t]
  \centering
  \begin{tabular}{r r r c c}
    \toprule
    Model ID & Input Vertices & Intersection Points & Ours' time & \cite{cherchi2020fast}' time \\
    \midrule
    356074   & 14K    & 38      & 0.02s  & 0.02s  \\
    199665   & 506K   & 1.2K    & 0.46s  & 0.45s  \\
    71377    & 502K   & 4.3K    & 0.60s  & 0.63s  \\
    622327   & 125K   & 34K   & 1.73s  & 1.67s  \\
    113906   & 24K    & 67K   & 1.35s  & 1.28s  \\
    1368052  & 1.2M   & 93K   & 3.51s  & 4.02s  \\
    247516   & 90K    & 106K  & 1.10s  & 1.11s   \\
    498460   & 95K    & 319K  & 8.94s  & 10.31s \\
    252786   & 355K   & 1.8M & 61.91s & 64.27s \\
    \bottomrule
  \end{tabular}
  \caption{In this table, we report the statistics of 9 "interesting" models from Thingi10k~\cite{Thingi10K} selected by \cite{lazard:hal-04907149}. For each model, we report the model's name, the number of input vertices (Input), the number of detected intersection points, the execution time of our detection and classification function (Ours' time), and the time of the reference function in \cite{cherchi2020fast} in seconds.}
  \label{tab:classification_times}
\end{table*}

In Table~\ref{tab:classification_times}, we report some numbers of a subset of "interesting" models selected by \cite{lazard:hal-04907149}, which varies from 38 to 1.8M detected intersections. For each model, we reported the model, the number of input vertices, the number of intersection points, the execution time of our tool and the one of \cite{cherchi2020fast}. As you can see, the execution times are also perfectly comparable for these models.

The results of this analysis clearly show that we are as robust and as fast as one of the most famous mesh Arrangements algorithms with a public implementation. Note that \cite{cherchi2020fast}'s algorithm is able to find and classify intersections in triangles represented exclusively with floating-point coordinates. Furthermore, their intersection detection module is fully integrated into their implementation and deeply dependent on their data structures. Therefore, the implementation of \cite{cherchi2020fast} cannot be smoothly extended to additional numerical representations or easily integrated into new algorithms.
Instead, our tool already supports alternative numerical representations and can be easily extended with additional ones. Furthermore, our code is generic and can effortlessly be included in new pipelines. All of these features are guaranteed without sacrificing robustness or performance compared to the state-of-the-art algorithm.

\section{conclusion and future works}
\label{sec:conclusion}

In this paper, we have presented a robust, fast, and easy-to-use tool for detecting and classifying triangle intersections. The strengths of this algorithm, implemented in C++ as a templated and header-only library, are more than one. First, it works with three different numerical representations (floating-point, rational numbers, and implicit points) and can be easily extended with additional ones. Second, it is implemented in a generic way and can be easily included in new algorithms under development. Furthermore, from a careful testing and comparison phase with a state-of-the-art algorithm~\cite{cherchi2020fast}, the tool has been demonstrated to be equally fast and robust, but with the possibility of being easily integrated and extended, without dependencies on data structures or algorithms specific to implementations of specific projects.

We believe that this tool can be very useful for several sub-fields of the scientific community (and not only) in Computer Graphics areas. For this reason, our implementation is available under the MIT Open-Source license in the following repository: \href{https://github.com/cg3hci/Fast-and-Robust-Tri-Intersections-with-different-Num-Reprs/tree/main}{\texttt{\textbf{LINK}}}.

As future developments, we are already working to support new numerical representation. Furthermore, we are already optimizing the code to make it more performant in terms of resources and execution times. 

In a subsequent release of the code, we plan to optimize the pipeline in order to reduce the number of geometric predicates (Orient2D and Orient3D) invoked. In fact, the bottleneck of our code lies in the number of predicate calls, which, after careful analysis of the problem from a theoretical and combinatorial point of view, \rev{we hope to be able to significantly reduce the predicate calls}, thus significantly reducing execution times.

Finally, since our goal is to help the community in their prototyping and development phases of new algorithms, we are open to help and code optimization from anyone who wants to contribute.

\section{Acknowledgments}
The authors deeply thank Marco Livesu, for the precious technical and scientific support to this work.
The authors gratefully acknowledge the support to their research by project ``FIATLUCS - Favorire l'Inclusione e l'Accessibilità per i Trasferimenti Locali Urbani a Cagliari e Sobborghi'' funded by the PNRR RAISE Liguria, Spoke 01, CUP: F23C24000240006. 

%
%


\bibliographystyle{alpha}
\bibliography{biblio}
\end{document}